\documentclass[aps,prl,onecolumn]{revtex4}
\usepackage{amsmath}
\usepackage{graphicx}
\usepackage{amsfonts}
\usepackage{epstopdf}
\usepackage{float}
\begin{document}
\title{Nature of light correlations in ghost imaging}
\author{Sammy Ragy}
\author{Gerardo Adesso}
\affiliation{School of Mathematical Sciences, The University of Nottingham, University Park, Nottingham NG7 2RD, United Kingdom}

\begin{abstract}
{\bf We investigate the nature of correlations in Gaussian light sources used for ghost imaging. We adopt methods from quantum information theory to distinguish genuinely quantum from classical correlations. Combining a microscopic analysis of speckle-speckle correlations with an effective coarse-grained description of the beams, we show that quantum correlations exist even in `classical'-like thermal light sources, and appear relevant for the implementation of ghost imaging in the regime of low illumination. We further demonstrate that the total correlations in the thermal source beams effectively determine the quality of the imaging, as quantified by the signal-to-noise ratio.}
\end{abstract}
\date{April 27, 2012}
\maketitle
\begin{figure}[ht]
\centering
\includegraphics[width=8cm]{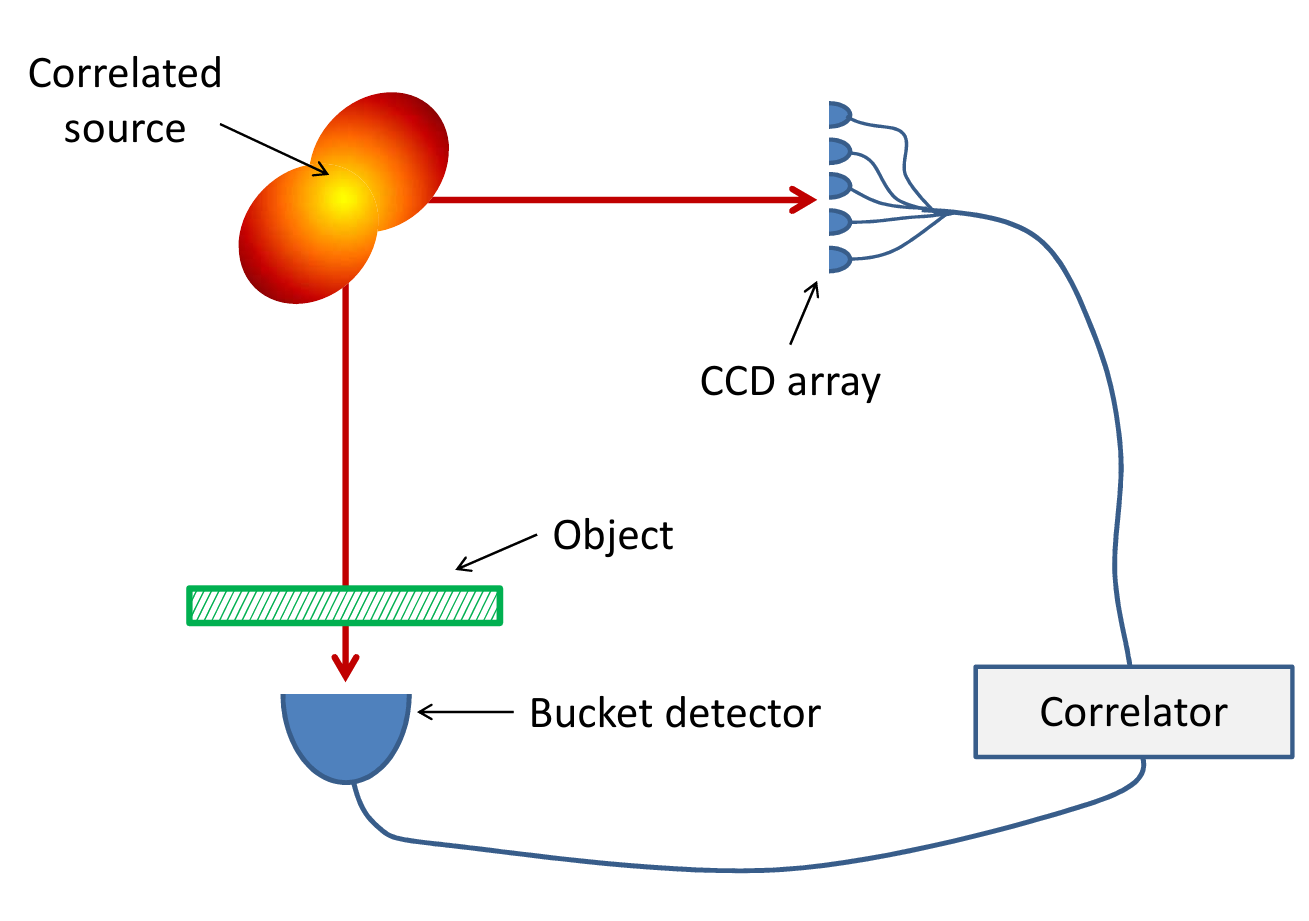}
\caption{Diagrammatic representation of a ghost imaging scheme.}
\label{scheme}
\end{figure}
Ghost imaging \cite{Sergienko,book} is an imaging modality configured as in figure~\ref{scheme}. Behind the object to be imaged, a bucket detector is placed. This detector yields no spatial information about the object but merely informs of the net photon count or intensity of the light impinging upon it. On its own it cannot provide an image of the object. In the second branch of the scheme, however, there is a spatially resolving detector (e.g. a CCD array); in this case, the light impinging upon this detector never passes through the object so it also cannot, on its own, yield an image of the object. In order to obtain an image, it is necessary to correlate the outputs of these detectors. Additionally, we need a source of correlated light to begin with, to provide the illumination; it is the exact nature of this source which has been the cause of controversy.

The first demonstration of ghost imaging was carried out by Pittman et al. in 1995 \cite{Pittman}. In their paper, they used entangled beams produced by spontaneous parametric down conversion (SPDC) to perform the imaging. Even then, they noted briefly at the end of their paper that it may be possible to adapt their technique to `classical' light---a proposition which sparked debate \cite{Abouraddy}. When this was experimentally achieved by Bennink et al. in 2002 \cite{Bennink}, part of the debate was resolved. Over the following years, even more theoretical and experimental papers ironed out the differences between ghost imaging with `quantum' and `classical' light \cite{Gatti, Bennink2, Brambilla}. Notably for our purposes, Gatti et al. \cite{Gatti2} proposed thermal-light ghost imaging which was experimentally achieved shortly afterwards \cite{Valencia, Ferri}. This prompted a debate on the big question whether the physical origin of thermal-light ghost imaging can be explained entirely using classical intensity correlations and to this day disagreement still persists \cite{Debate1,Debate2}.

Despite the large number of studies addressing practical questions such as signal-to-noise ratio (SNR), image contrast and acquisition time \cite{Gatti2,Erkmen,ShapiroSNR,Sullivan}, to our knowledge, there has been a lack of study on the specific decomposition of classical and quantum correlations (in their broadest sense) in the light source used for ghost imaging. We believe that a study from such an angle will elucidate how quantumness versus classicality manifest in the scheme.

One qualifier of `classicality' for ghost imaging has been that it can be accounted for by a semiclassical optics description---where light is treated classically, but the quantum nature of the detector is accounted for by consideration of the shot-noise \cite{Erkmen,Review}. Thermal-light ghost imaging can indeed afford such a description. However, an immediate issue arises: for two modes of Gaussian light between which any correlation exists at all, a nonvanishing part of the total correlations has  necessarily a genuinely quantum nature \cite{Discord,Discord2}, as captured for instance by the quantum discord \cite{Zurek,Vedral}. Since the correlated thermal light used for such experiments is indeed Gaussian, it warrants a study of the extent to which these quantum correlations feature in so-called `classical' ghost imaging.

To put this on a more rigorous footing, in quantum optics, `classical' light is taken to be light with a proper (non-negative and non-singular) Glauber $P$-representation, meaning it can accurately be represented by classical optical theory \cite{Mandel}. For two-mode Gaussian states in standard form \cite{ourreview}, classicality as qualified by the $P$-representation criterion coincides with separability, that is, absence of entanglement \cite{slater}. However, this classification does not correspond to the most general {\it informational} definition for the classicality of correlations in bipartite states. According to a recently established paradigm, rooted in quantum foundations and information theory, a state is classically correlated when it has vanishing quantum discord \cite{Zurek,Vedral}, i.e., when a projective measurement exists that once performed on a subsystem leaves the composite state undisturbed. In this respect, almost all unentangled states possess general quantum correlations as well \cite{ferraro}. Quantum correlations without and beyond entanglement are capturing the attention of a vast community of researchers, as those correlations have been linked to practical advantages in a number of noisy quantum communication and computation frameworks where entanglement is too fragile to be maintained \cite{dreview,merali}. In fact, on general grounds, a recent work has indicated that the sets of states corresponding to each definition of classicality (the optics one in terms of $P$-representation, and the informational one in terms of discord) are almost completely disjoint \cite{Matteo}. Since the ghost imaging scheme is entirely dependent on the correlations between each beam, one needs to apply special care when tagging any instance of that protocol as `classical' based on one perspective only. Considering that ghost imaging is a scheme which entirely depends upon the correlations between each arm, it seems likely that an analysis from the informational perspective of correlations may shed some light on the problem.

To better understand the debate, it is useful to divide thermal-light ghost imaging into two categories: lensed and lensless. At present, advocates for the quantum description of thermal-light ghost imaging propose that the ghost image formation in the lensless case cannot be explained semiclassically but can only be accounted for by non-local two-photon interference, a distinctly quantum phenomenon. It is the indistinguishability of different ways of triggering a joint detection which is crucial for this characterisation \cite{Debate1,Scarcelli}. More so, there have been claims that the non-local interference picture of ghost imaging is necessary to explain the results of some recent experiments. These include ghost imaging with speckle-free light \cite{Sun} (although this idea of speckle-free light has met some resistance \cite{Debate2}) and also an experiment on turbulence-immune ghost imaging \cite{Turb1}, but due to the general contentiousness of ghost imaging this too has come under scrutiny \cite{Debate2,Turb2}.

On the other hand, the lensed picture is much less controversial. Even in terms of the two-photon interference picture, it is considered as a ``man-made'' local correlation; the propagation characteristics induced by the lens (i.e. a Fourier transform), mean that each mode in the source plane is mapped to a unique position in the detection planes and a local, semiclassical picture can be used. A detailed description of this can be found in \cite{Review2}. Nevertheless, this type of ghost imaging still warrants an examination from our perspective due to the aforementioned presence of quantum correlations in \textit{all} correlated Gaussian states.

In this paper, we quantitatively compare the correlations in the source Gaussian light for lensed ghost imaging to a specific figure of merit, the SNR. In the case of thermal-light ghost imaging, we find that there is a universal quasi-linear relation between the \textit{total} correlations (the sum of quantum and classical) and the SNR. Furthermore, our analysis reveals two regimes: one---for low illumination---in which quantum effects dominate (in the sense that quantum discord takes up most of the total correlations), and another---for high illumination---in which it is the classical side of correlations which is much stronger. Increasing the illumination intensity we thus witness a quantum-to-classical transition in the description of ghost imaging, all the time using unentangled thermal light, which is classified as entirely `classical' by optics standards. In the limit of very high illumination, thermal light and entangled sources yield the same performance for ghost imaging, and we find that they are characterised by the same, mostly classical correlations (although we re-emphasise that the quantum component of correlations is only exactly zero for completely uncorrelated light). We also examine the behaviour of correlations in near-field, lensless ghost imaging, removing any dependence on speckles from our calculations. We find that both lensed and lensless ghost imaging behave in a similar manner.

\section{Results}

\subsection{Overview}
\begin{figure}[b]
\centering
\includegraphics[width=10cm]{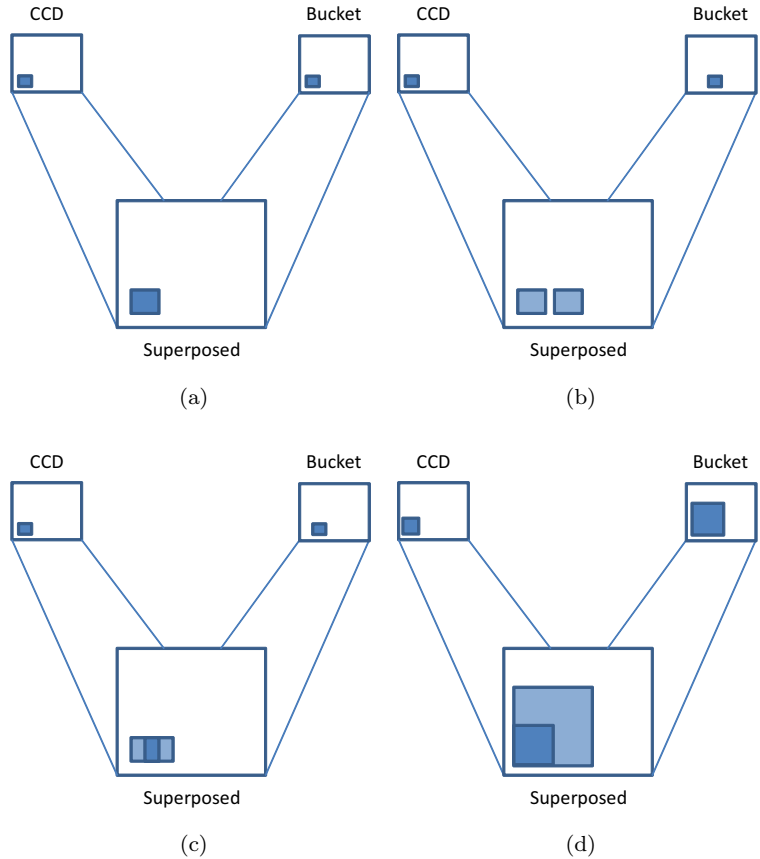}
\caption{Conceptual diagrams indicating the areas we are considering on the CCD and bucket planes and a superposition of both. For part (a) we are considering exactly matched areas on either plane, for part (b) we are considering entirely disjoint areas. Part (c) has two areas of equal magnitude, half overlapping. The correlations between these are $\frac{1}{2}$ those for an individual speckle within the overlapping area. Part (d) considers one area $\frac{1}{4}$ the size of the other, completely enclosed by it on the superposed picture. This case notably, is most like the comparison of pixel and bucket detector. The effective modes also have $\frac{1}{2}$ the cross-correlations between them as for any individual speckles within the overlap area.}
\label{Planes}
\end{figure}

Before delving into our results, it is necessary to clarify which variables influence the light correlations and the SNR simultaneously for our lensed picture. To begin we have the speckle-count per pixel, $M$ of the spatially resolving detector. When we refer to `speckle', we mean the spatial intensity fluctuations projected upon the detectors by a spatially incoherent beam \cite{Speckle}. The second parameter of interest is the illumination $I=M \mu$, where $\mu$ denotes the average photon count per mode---which for our lensed formulation corresponds to the photon count per speckle. The final parameter of interest is the pixel count of the image, $R$, which informs us over how many modes the bucket detector integrates. More details on this formulation of ghost imaging can be found in \cite{Brida}, from which we have taken the SNR expression. Out of several possible choices \cite{Brida,ShapiroSNR}, we choose to adopt the SNR corresponding to an imaging function which utilises the covariance of the intensities in each beam. The explicit expressions of the SNR for thermal-light and SPDC-entangled light sources are provided in the Methods section.

In the lensed ghost imaging, we have pairwise correlated speckles. Each pair of correlated modes in the source-plane is mapped onto a unique position in each of the CCD and bucket planes. As such, any speckle is correlated to one---and only one---speckle on the opposite plane. This is a very typical setup for the observation of so-called `classical' ghost imaging \cite{Valencia,Ferri,Brida}.

We shall look at the character of two different types of correlations: the individual (microscopic) speckle-speckle correlations for each pair of correlated speckles, and a  `coarse-grained' view which averages over the entire bucket detector for each pixel of the spatially resolving one. This latter view, which allows us to adopt an effective two-mode description of the problem, gives an impression of the correlations involved in shaping each pixel of the ghost image with a close regard to how the ghost imaging scheme is actually constructed.
There are several note-worthy aspects of the coarse-grained formalism which we shall elucidate diagrammatically; a full description of the derivation can be found in the Methods section.

We specialise here our qualitative picture to thermal-light lensed imaging. We define effective operators for an area on each plane, $\hat{a}_{A_1}$ and $\hat{a}_{A_2}$. It turns out that for any given area, $A$, on either plane, the autocorrelations  $\langle\hat{a}^\dagger_{A}\hat{a}_{A}\rangle$ are equal to those for any individual speckle within the area. If we look at any two matched areas on the CCD and bucket planes (figure~\ref{Planes}(a)), then the cross-correlations, $\langle\hat{a}^\dagger_{A_1}\hat{a}_{A_2}\rangle$, are again exactly equal to the cross-correlations of the \textit{individual} paired speckles (all speckles are assumed to have identical statistics). Similarly if we look at any two \textit{unpaired} areas, in figure~\ref{Planes}(b) then we get no correlations at all, which is exactly as we would expect. In this sense the coarse-graining proves to be a sensible method of averaging.
Better yet, if we look at partially overlapping regions and regions of disparate area, the cross-correlations depend on the ratio of the overlap to total area. It scales as $\frac{A_{overlap}}{\sqrt{A_1 A_2}}$. This accounts for the cases of figure~\ref{Planes}(c) and (d). Panel (d) in particular is of relevance for ghost imaging; in this case, we find that the cross-correlations scale as $\sqrt{\frac{A_{overlap}}{A_2}}$, or equivalently $\sqrt{\frac{A_{\text{pixel}}}{A_\text{bucket}}}=\frac{1}{\sqrt{R}}$.

In the following, quantum correlations ${\cal Q}$ will be quantified by the Gaussian quantum discord \cite{Zurek,Discord,Discord2}, classical correlations ${\cal C}$ by the complementary one-way correlation measure introduced in \cite{Vedral}, and total correlations ${\cal T}$ by the conventional quantum mutual information, which coincides with the sum of the two.

\subsection{Speckle-speckle correlations}
\begin{figure}[H]
\centering
\includegraphics[width=16cm]{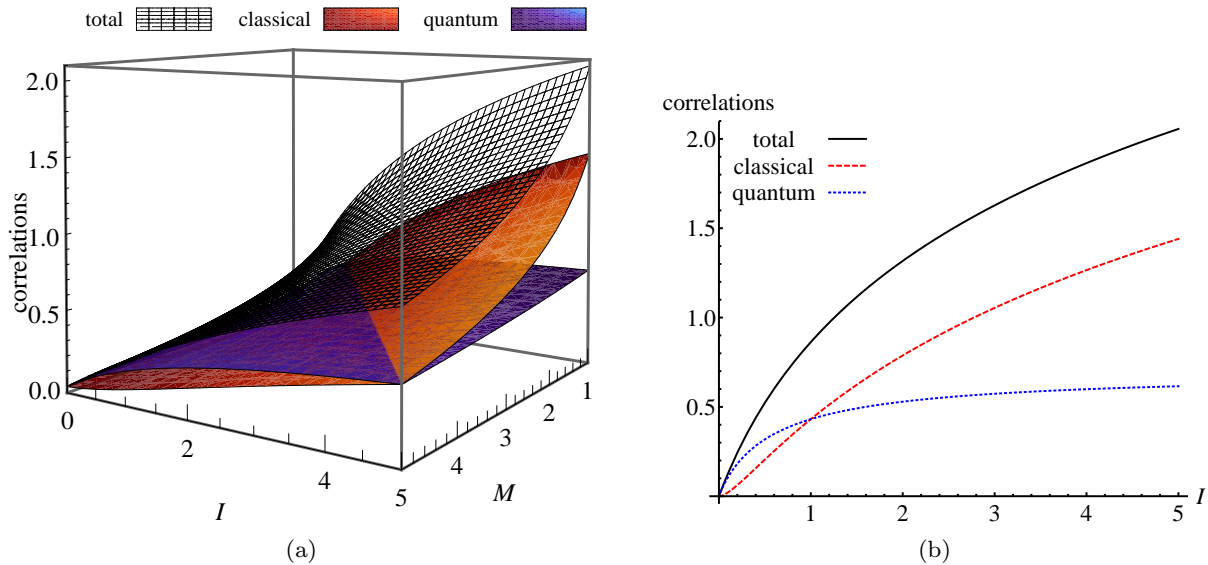}
\caption{Speckle-speckle correlations in thermal-light lensed ghost imaging, plotted as a function of the illumination $I$ and the speckle-count $M$. In (b) a detail at $M$=1 is shown. Notice the intersection between classical and quantum correlations at $M=I$. Classical ones dominate for $I>M$, while quantum ones are relevant for $I<M$.}
\label{single}
\end{figure}

Looking at the speckle-speckle correlations, we get our first hints that even with non-entangled thermal light, the role of quantum correlations cannot be ignored. In figure~\ref{single}, we clearly see that for low illumination $I$---or alternatively, high speckle-count per pixel $M$---quantum correlations can actually \textit{exceed} classical ones. This stems from the fact that for individual pairs, the correlations only depend on the expected photon count $\mu=\frac{I}{M}$. When $\mu<1$ the speckle-speckle quantum correlations dominate over the classical ones, ${\cal Q} > {\cal C}$. Entanglement is never present in the considered light source, yet our study reveals a definite non-classical nature of such light, manifested in the correlations between individual pairs of  speckle modes in the low illumination regime.

\subsection{Coarse-grained correlations}
The results we find for the coarse-grained correlations are even more striking. In the effective two-mode description, we find it convenient, for the sake of the discussion below, to normalise the correlations multiplying them by a factor $\sqrt{\frac{R}{2}}$ (see Methods for a justification). In the following, we refer to the normalised total, quantum, and classical coarse-grained correlations by $\tilde{\cal T}$, $\tilde{\cal Q}$, and $\tilde{\cal C}$, respectively. Our aim is to compare their behaviour with the SNR, which quantifies the quality of the imaging. In this way we can attempt to identify which aspects of the correlations in the source beams can be regarded as key resources for the protocol.

\begin{figure}[b]
\includegraphics[width=16cm]{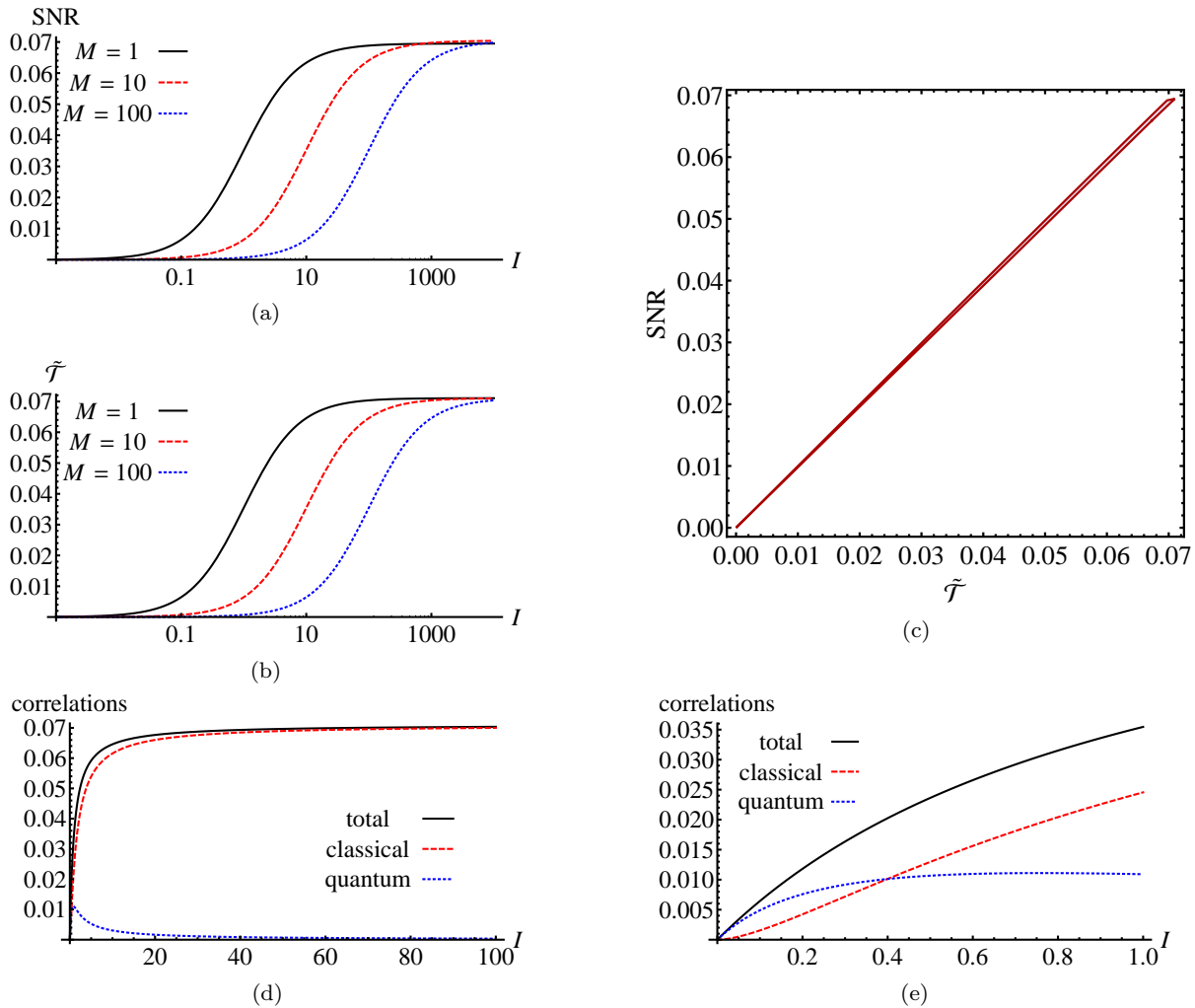}
\caption{Coarse-grained analysis for thermal-light ghost imaging (with pixel count $R=100$). Panels [(a)-(b)]: Log-linear plots of (a) Signal-to-noise ratio (SNR) and (b) total correlations $\tilde{\cal T}$ for increasing illumination $I$. In panel (c) we display the tiny region filled by SNR versus $\tilde{\cal T}$ with varying $M \in [1,1000]$ and $I \in [0,1000]$; the quasi-linear region is essentially invariant upon variations of $R$ as well. Panels [(d)-(e)] display the decomposition of coarse-grained correlations, for $M=1$. Looking at (d) we can see that the total correlations are bounded, and at high illuminations they are mostly classical. Zooming in the low-illumination regime near the origin, though, we see in (e) that there is an interval where quantum correlations still exceed classical ones. All the correlations are normalised by $\sqrt{R/2}$.}
\label{SNR}
\end{figure}

To begin, we first plot the {\it total} normalised coarse-grained correlations $\tilde{\cal T}$ and the SNR on separate graphs in figure~\ref{SNR}. Despite the very different (physical and mathematical) nature of the two quantities under scrutiny, it is immediately noticeable that they have a very similar form, a fact which is even more evident when we observe a parametric plot of the SNR versus $\tilde{\cal T}$ in figure~\ref{SNR}(c). By varying $I$, $M$, and $R$ in their region of interest (in particular keeping $R \gg 1$, which means nontrivial imaging), we find that the SNR always exhibits a quasi-linear dependence on the total correlations with slope $\approx 1$. The relation is not exactly linear, yet the discrepancy between SNR and normalised total correlations stays smaller  than one percent in the relevant parameter regime.  This indicates that for \textit{any} of the parameters that affect the light correlations, the SNR is affected in exactly the same way.  The small difference may reflect the fact that there are properties of the detector and object which affect the SNR, but do not  influence the correlations at the source.
Rigorously, recalling $\mu=I/M$, and using the formulas provided in the Methods, we have
\begin{equation}\label{eqlimit}
\frac{\rm SNR}{\tilde{\cal T}} \underset{\mu \rightarrow \infty}{\longrightarrow} \frac{\sqrt{1/(2R+1)}}{\sqrt{R/2} \ln[R/(R - 1)]} \underset{R \rightarrow \infty}{\longrightarrow} 1\,.
\end{equation}

Our analysis thus reveals that the joint contribution of quantum {\it and} classical correlations can actually be used as a predictor for the performance, as measured by the SNR, of ghost imaging with `classical', thermal-light sources. It is now of interest to look at how the total correlations decompose specifically for our coarse-grained description. In figure~\ref{SNR}(d), we see that for high values of illumination, the quantum correlations $\tilde{\cal Q}$ tend to $0$. As such, total correlations are almost all classical (and, like the SNR, they are bounded from above). We do, however, still see a region near the origin [figure~\ref{SNR}(e)] where the coarse-grained quantum correlations exceed the classical ones, in agreement with the speckle-speckle analysis of figure~\ref{single}.

\begin{figure}[t]
\centering
\includegraphics[width=16cm]{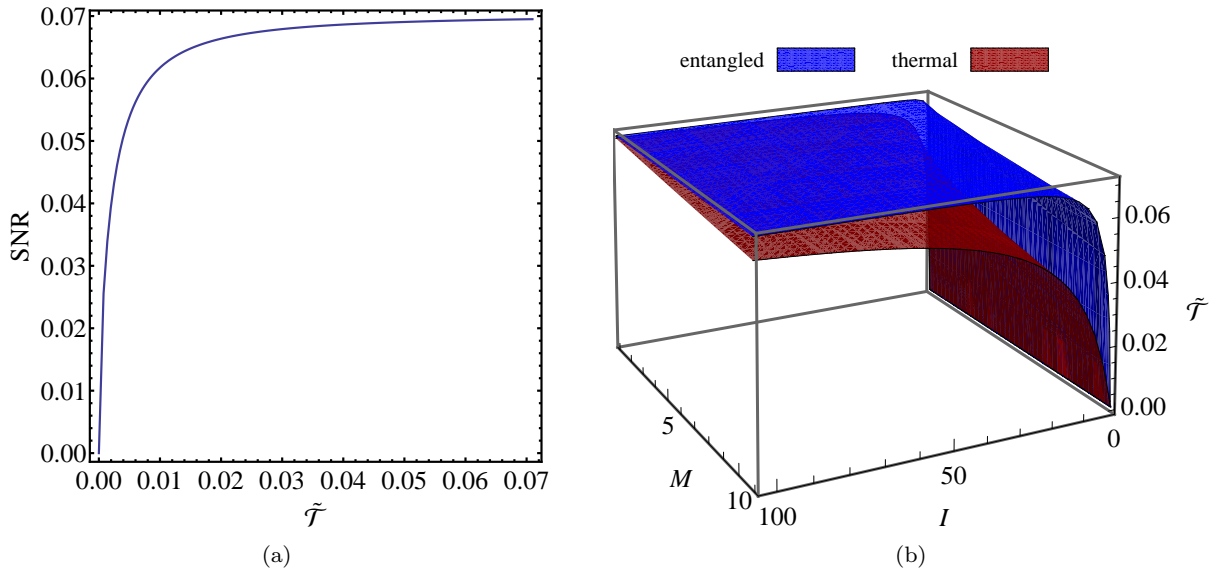}
\caption{(a) Coarse-grained analysis for ghost imaging using entangled light produced by SPDC (with pixel count $R=100$): Plot of SNR versus total correlations for $M=1$ and $I\in[0,1000]$; compare with the corresponding thermal-light case, figure \ref{SNR}(c). Panel (b) depicts a comparison between total coarse-grained correlations ${\tilde {\cal T}}$ for the cases of thermal and entangled light sources (with $R=100$) as a function of the illumination $I$; it is shown that they share a common limit in the regime of high illumination [see Eq.~(\ref{eqlimit})], although in this setup SPDC-entangled light is always more correlated than thermal-light.
All the correlations are normalised by $\sqrt{R/2}$.}
\label{QvsC}
\end{figure}

\subsection{Extending the picture}
We have comprehensively unveiled the presence and role of genuinely quantum correlations in thermal-light, `classical'-like ghost imaging. We can also extend our study to the case of ghost imaging with entangled light sources produced by SPDC. In this case, it has already been pointed out in \cite{ShapiroSNR} that the entanglement between the individual modes acts as an extra resource and can lead to an increase in the SNR compared to the thermal-light case. In our formalism, we find that for entangled light the SNR (see Methods) grows much faster than linearly as a function of the coarse-grained total correlations $\tilde{\cal T}$, as shown in figure~\ref{QvsC}(a). However, in the limit of very high illumination, the coarse-grained total correlations for both thermal-light and entangled cases converge to the same asymptotic maximum value, which is an intuitively expected result \cite{Erkmen}. Similarly, the SNR for both cases converges to the same upper limit in the regime of high illumination, given by ${\rm SNR}_{\mu \rightarrow \infty} = \sqrt{M/[6+M(2R+1)]}$ \cite{Brida}. Specifically, we find that Eq.~(\ref{eqlimit}) holds exactly for entangled as well as thermal light. When we plot the (normalised, coarse-grained) total correlations for thermal-light versus SPDC-entangled sources on the same graph, as in figure \ref{QvsC}(b), it is evident though that in the case of entangled light the correlations are always higher for given finite values of the parameters $I$, $M$ and $R$.

So far we have made quantitative comparisons for lensed ghost imaging only, however it turns out that the correlations for near-field, lensless ghost imaging also scale in precisely the same way. Instead of considering speckle-count per pixel and pixel-count over the image, we simply calculate the correlations between pixel area and bucket detector area. The qualitative results of our previous description then still hold, and we have included detailed calculations in the Methods section.

\section{Discussion}

We have analysed the nature of correlations in Gaussian light sources used for ghost imaging from a quantum informational perspective, combining a microscopic with an effective coarse-grained description. We have found that even so-called `classical' thermal light contains nonzero genuinely quantum correlations---as measured by the quantum discord \cite{Zurek,Discord}---whose contribution to the performance of ghost imaging schemes has been assessed. Since the entire scheme is dependent upon the correlations between the arms, the quasi-linearity between the total correlations and SNR strongly suggests that the total correlations may be acting as the essential operative resource.

The dominant strength of quantum correlations in low-illumination regimes has an immediate physical explanation. In the limit of low illumination, there are few photons per mode and this is a regime where the quantum behaviour of light becomes very apparent. In fact, in a recent review, the non-local description for Gaussian thermal light is formulated with the explicit assumption of low photon-flux \cite{Review} and in the original paper expounding the non-local picture \cite{Scarcelli}, the adopted model of thermal light implicitly assumes a photon-counting regime, which is an equivalent criterion to low photon-flux. As elucidated by the results of this paper, it is likely no mere coincidence that such an assumption need be made, but rather a consequence of the fact that the quantum component of correlations available for detection by our scheme vanishes in the limit of high illumination.

This reveals an interesting feature associated to the coarse-grained formalism put forward here. It actually indicates how the quantum nature of the light source becomes quenched as we diverge from the photon-counting regime and enter the classical limit of intensity correlations. For these high illuminations, the quantum correlations available for detection by our scheme tend to zero, and the physical model of the scheme does not require a quantum description of the light to be accurate. The small-scale features of the speckle-speckle correlations cannot be detected by the way the light is averaged over the bucket detector. This is {\em an unusually clear look into how a system can transition from a quantum regime into a classical one}.

We have briefly extended our analysis to the less controversial case of SPDC-entangled light sources. It has been observed in previous work \cite{Erkmen} that for high brightness the results of the ghost image formed with `classical'-like thermal  light are ``excellent approximations for the quantum [entangled] case'' and we find indeed that in this limit, the (coarse-grained) total correlations in the sources too approximate each other closely.

Our calculations, as explicated in the Methods section, reveal furthermore that lensed and lensless ghost imaging behave in a similar manner to each other. In the recent debate, which is well characterised by a comment from Shih \cite{Debate1} with a rebuttal from Shapiro et al. \cite{Debate2}, we find our results support the latter's arguments that near-field lensless imaging \textit{can} afford a classical interpretation. In high illuminations the quantum correlations are suppressed. Ultimately, this shows that the geometry of the system has a limited role in determining its classicality or quantumness, whereas the illumination conditions are a clearly more significant factor.

We finally mention that quantum discord has been very recently measured experimentally in Gaussian states of light \cite{ulrik,milegu,blandino}. We believe our predictions on the role of quantum versus classical---and total---correlations in ghost imaging with thermal Gaussian light could be tested with current optical technology.

\section{Methods}
\subsection{Correlations in Gaussian states}
A Gaussian state is any state which can be represented by a Gaussian Wigner distribution in phase space. Ranging from thermal to squeezed states, they are convenient not only for their experimental ubiquity and robustness, but also for their ease of use for calculations.

Gaussian states are entirely characterised by their first and second statistical moments. For this paper, we deal with zero-mean states; as such, the relevant object for our computations is the two-mode covariance matrix given by $\sigma_{ij}=\text{Tr}[ \hat{\rho} (\hat{R}_i\hat{R}_j+\hat{R}_j\hat{R}_i)]$. Here $\hat{R}_{i,j}$ are the elements of our quadrature vector given by $\hat{R}=(\hat{q}_1,\hat{p}_1,\hat{q}_2,\hat{p}_2)^T$ \cite{ourreview}.

From this, we can fully determine the information required for computing the correlations between two modes. The measure of quantum correlations which we choose is the quantum discord \cite{Zurek,Vedral}, which is defined for bipartite states as the difference between two classically identical definitions of mutual information. When the difference is non-zero, we then necessarily have quantum correlations in the state.

Any two-mode covariance matrix can be written in the form
$$
\sigma_{12}=\begin{pmatrix}
a&0&c&0\\0&a&0&d\\c&0&b&0\\0&d&0&b
\end{pmatrix}.
$$
Subsequently, we can write the quantum discord, the classical correlations and the mutual information, respectively, as
\begin{align*}
\mathcal{Q}(\sigma_{12})&=f(b)-f(\nu_+)-f(\nu_-)+\inf f(\sqrt{\det\epsilon}),\\
\mathcal{C}(\sigma_{12})&=f(a)-\inf f(\sqrt{\epsilon}),\\
\mathcal{T}(\sigma_{12})&=f(a)+f(b)-f(\nu_+)-f(\nu_-),
\end{align*}
where $f(x)=\frac{x+1}{2}\ln(\frac{x+1}{2}) +\frac{x-1}{2}\ln(\frac{x-1}{2})$, $\nu_\pm$ are the symplectic eigenvalues of the covariance matrix \cite{ourreview} and $\epsilon$ is the conditional covariance matrix of subsystem $1$ after an optimal Gaussian measurement has been performed on subsystem $2$. The full details on the interpretation of these quantities and on of how to perform these calculations can be found in \cite{Discord}.
\subsection{Lensed Ghost Imaging and SNR}
For this section we have utilised the model of Brida et al. \cite{Brida} and a detailed analysis can be found in their paper. In the lensed setup, modes with particular values of transverse vector $\kappa$ are projected onto unique transverse coordinates $\rho$ in the imaging plane.  We assume that the modes corresponding to each value of $\kappa$ have identical statistics. Then, for the thermal-light case we can model the speckled beam as a set of independent modes $\hat{a}_i$. Performing the beam splitter transformation then leads to new operators $\hat{b}_{1,i}=(\hat{a}_i+\hat{a}_{vac,i})/\sqrt{2}$ and $\hat{b}_{2,i}=(\hat{a}_i-\hat{a}_{vac,i})/\sqrt{2}$. The statistics for each of these beams individually are identical.

The SPDC-entangled case requires a very similar calculation. The difference is that our modes are generated by the well-known equations $\hat{b}_{1,i}=U \hat{a}_{1}({q_i}) + V \hat{a}^\dagger_{2}(-q_i)$ and $\hat{b}_{2,i}=U \hat{a}_{2}(q_i) + V \hat{a}^\dagger_{1}(-q_i)$, where $U^2-V^2=1$.

It is now useful to consider the signal-to-noise ratio of ghost imaging. The SNR is defined as
$$
\text{SNR}= \frac{|\langle S_{in}-S_{out}\rangle |}{\sqrt{\text{var}(S_{in}-S_{out})}}
$$
where our object is a binary amplitude mask; that is, we either have full transmission or full occlusion of the incoming light. $S_{in}$ corresponds to being in the object profile (full transmission), and $S_{out}$ the opposite. Here $S$ is whichever imaging function we choose.

As mentioned in the main text, for our analysis we are interested in three parameters. The first is the number of speckles, $M$, per pixel of the spatially resolving detector. The second, is the illumination, $I=M\mu$, where $\mu$ is our mean photon number per spatiotemporal mode, and the last is the number of pixels in our ghost image $R$.

Choosing to take our imaging function $S$ to be the correlation function of intensity fluctuations (the covariance) we have $S(\rho)=\langle\mathbb{N}_1-\langle\mathbb{N}_1\rangle\rangle\langle N_2(\rho)-\langle N_2(\rho)\rangle\rangle$. The spatial dependence of $N_2$ refers to the position in the spatially-resolving arm. Furthermore, $\mathbb{N}_1$ denotes the net photon count on the bucket detector. To clarify, we have $\mathbb{N}_1=\sum_{j=1}^R N_{1,j}$, where $N_{1,j}$, refers to our photon count on the $j^{th}$ spatial resolution cell. In turn we can decompose each of our $N_1$, $N_2$ into the sum of individual modes (speckles) such that $N_i=\sum_{j=1}^M n_{i,j}$ where $i=1,2$. This quantity tells us the number of speckles per pixel.

We then have \cite{Brida}
\begin{eqnarray}\label{eqsnrs}
\text{SNR}_{\rm thermal} &=& \frac{\mu  \sqrt{M}}{\sqrt{\mu ^2 (2 M R+M+6)+4 \mu  (M R+1)+2 M R+1}}\,,\\
\text{SNR}_{\rm entangled} &=& \frac{\sqrt{\mu  (\mu +1) M}}{\sqrt{\mu ^2 (2 M R+M+6)+\mu  (2 M R+M+6)+1}}\,,
\end{eqnarray}
for thermal-light and SPDC-entangled light sources, respectively.

\subsection{Coarse-grained description for lensed imaging}
As outlined earlier, the SNR depends on an imaging function of the form $S(x)=\langle\mathbb{N}_1-\langle\mathbb{N}_1\rangle\rangle\langle N_2(x)-\langle N_2(x)\rangle\rangle$, which takes into account multiple modes. Whilst it is easy to generate multimode covariance matrices, the problem of calculating the discord from these matrices is quite intractable. As such, we proceed to define an effective two-mode matrix. The key to this is to produce averaged operators which, under the expectation value, behave like usual single-mode operators.

To wit, starting with a pixel which collects $M$ modes and noting that $N_i=\sum_{j=1}^M n_{i,j}=\sum_{j=1}^M \hat{b}^\dagger_{i,j} \hat{b}_{i,j}$ (for $i=1,2$), we wish to find mode operators $\hat{c}_i$ for each pixel, such that $\langle\hat{c}_i^\dagger\hat{c}_i\rangle=\langle\sum_{j=1}^M \hat{b}^\dagger_{i,j}\hat{b}_{i,j}\rangle$. In order to do this, we need to explicitly consider the spatial dependence of the modes at the detection planes. Doing so, we can say $\hat{c}_i=\sum_{j=1}^M \hat{b}_{i,j}\delta(\rho-\rho_j)$ where $\rho_j$ corresponds to the position of the $j^{\text{th}}$ mode on the orthogonal plane. Note that we do not consider the effects of the transmission mask on light propagation in this particular calculation.

These operators then behave as we desire, but for one very crucial part: the commutation relations. This is easily remedied though. If we define our effective operators in the following manner: $\hat{d}_i=\frac{1}{\sqrt{M}}\hat{c}_i$, then we find $\langle[\hat{d},\hat{d}^\dagger ]\rangle=1$. Performing similar steps for pixels collecting $R$ modes, and then transforming our covariance matrices from the mode operator to the quadrature basis, enables us to obtain the effective `coarse-grained' covariance matrices given by
\begin{gather}
\begin{pmatrix}1+2\mu & 0 & \frac{2\mu}{\sqrt{R}} &0\\0 & 1+2\mu & 0 & \frac{2\mu}{\sqrt{R}}\\ \frac{2\mu}{\sqrt{R}}&0&1+2\mu & 0 \\ 0& \frac{2\mu}{\sqrt{R}}&0&1+2\mu\
\end{pmatrix}\,, \quad \mbox{and} \quad
\begin{pmatrix}1+2\mu & 0 & \frac{2\sqrt{\mu+\mu^2}}{\sqrt{R}} &0\\0 & 1+2\mu & 0 & \frac{2\sqrt{\mu+\mu^2}}{\sqrt{R}}\\ \frac{2\sqrt{\mu+\mu^2}}{\sqrt{R}}&0&1+2\mu & 0 \\ 0& \frac{2\sqrt{\mu+\mu^2}}{\sqrt{R}}&0&1+2\mu
\end{pmatrix}\,,
\label{cms}\end{gather}
for thermal-light and SPDC-entangled light cases, respectively.

Notice that the off-diagonal blocks of the covariance matrices, which encode the intermodal correlations, appear scaled by a factor $\propto 1/\sqrt R$ as a consequence of the averaging applied to preserve the commutation relations. This makes e.g.~the covariance $\langle \hat{a}_{A_1}^\dagger \hat{a}_{A_2}\rangle$ sensitive to changes in one of the areas, which should not be, as an increase in, say, $A_2$ for a fixed $A_1\leq A_2$ implies no loss of any physical correlation. Therefore, to make sure we measure genuine phase-insensitive correlations between the beams in the effective two-mode description, we can {\it a posteriori} renormalise classical, quantum, and total correlations, evaluated on the  matrices of Eq.~(\ref{cms}), multiplying them by a factor proportional to $\sqrt R$. The specific choice $\sqrt{R/2}$ is dictated by mere convenience, as it makes the slope in the SNR versus total correlations relation converge exactly to $1$ [see Fig.~\ref{SNR}(c) and Eq.~(\ref{eqlimit})].
Needless to say, the quasi-linear interdependence existing between the two quantities for thermal light, as well as all the results derived in the main text and illustrated in figures \ref{SNR} and \ref{QvsC}, are obviously not qualitatively affected (albeit for quantitative rescaling) by any specific normalisation procedure implemented on the correlations.


\subsection{Coarse-grained description for lensless ghost imaging}
The above formalism for lensed ghost imaging considers a simple one-to-one mapping from the operators in the momentum-basis at the source plane, to the position dependent ones at the detection planes. However, we can generalise our calculations to other sorts of propagation. For the purposes of lensless ghost imaging, it is conventional to consider free-space propagation in the paraxial approximation. In this case, it is possible to characterise the field at transverse point $\rho_j$ in plane $z_j$ by $\hat{E}(\rho_j,z_j;\kappa)\propto\sum_\kappa g(\rho_j,z_j;\kappa) \hat{a}(\kappa)$ for $j=1,2$, where $\hat{a}(\kappa)$ are our source-plane operators and $g(\rho_j,z_j;\kappa)$ is the Green's function describing the propagation of the field to the point with transverse coordinates $\rho_j$ on the detection plane \cite{Rubin, Review2}. The form of $g(\rho_j,z_j;\kappa)$ in the quasimonochromatic case is \cite{Rubin}:

$$
g(\rho_j,z_j;\kappa)=  \frac{-ik_0 e^{i k_0 z_j}}{2\pi z_j}\int d \rho_s e^{i \frac{k_0}{2 z_j}|\rho_j-\rho_s|^2} e^{-i\kappa\cdot\rho_s} .
$$
We can then calculate the the auto-correlations and cross-correlations and from this, our effective covariance matrix. To begin, we note that in the quasimonochromatic, paraxial approximation, $[\hat{E}_i(\rho_i,z_j),\hat{E}_j^\dagger(\rho_j,z_j)]=\delta(\rho_i-\rho_j)\delta_{ij}$, where $\rho$ are the coordinates in the transverse plane \cite{Erkmen, Shapiro}. As such, these behave under the expectation value as though they were the usual single mode operators $\hat{a}$. Henceforth, we will assume that we are always considering correlations on planes at equal distances from the source as is conventional for ghost imaging setups. For individual modes or correlated pairs:

\begin{align}
\langle\hat{E}^\dagger(\rho_i,z)\hat{E}(\rho_j,z) \rangle&\propto\langle\sum_{\kappa \kappa'}g^*(\rho_i,z;\kappa)g(\rho_j,z;\kappa')\hat{a}^\dagger(\kappa)\hat{a}(\kappa') \rangle \nonumber\\
&=\sum_{\kappa \kappa'}\langle g^*(\rho_i,z;\kappa)g(\rho_j,z;\kappa')\rangle\langle\hat{a}^\dagger(\kappa)\hat{a}(\kappa') \rangle \nonumber\\
&=\sum_\kappa \langle \left|\frac{k_0 e^{i k_0 z}}{2\pi z}\right|^2\int d\rho_s e^{-i \frac{k_0}{2 z}|\rho_i-\rho_s|^2} e^{i\kappa\cdot\rho_s} \int d\rho_s' e^{i \frac{k_0}{2 z}|\rho_j-\rho_s'|^2} e^{-i\kappa\cdot\rho_s'} \rangle\langle\hat{a}^\dagger(\kappa)\hat{a}(\kappa)\rangle \nonumber\\
&=\left(\frac{k_0}{2\pi z}\right)^2\langle\hat{a}^\dagger(\kappa)\hat{a}(\kappa)\rangle\sum_\kappa \langle \int d\rho_s d\rho_s' e^{i \frac{k_0}{2z} |\rho_j-\rho_s'|^2-|\rho_i-\rho_s|^2} e^{i \kappa \cdot (\rho_s-\rho_s')} \rangle \nonumber\\
&\propto\left(\frac{k_0}{2\pi z}\right)^2\langle\hat{a}^\dagger(\kappa)\hat{a}(\kappa)\rangle\langle \int d\rho_s e^{i \frac{k_0}{2z} |\rho_j-\rho_s|^2-|\rho_i-\rho_s|^2} \rangle \nonumber \\
&=\left(\frac{k_0}{2\pi z}\right)^2\langle\hat{a}^\dagger(\kappa)\hat{a}(\kappa)\rangle e^{i\frac{k_0}{2 z}(\rho_j^2-\rho_i^2)}\int d\rho_s e^{i \frac{k_0}{z}\rho_s \cdot (\rho_j-\rho_i)} \nonumber \\
&\approx \langle\hat{a}^\dagger(\kappa)\hat{a}(\kappa)\rangle \delta(\rho_j-\rho_i). \nonumber
\end{align}
For the calculation, we have assumed a large disk-like source and a large number of modes. These are standard assumptions made in the derivation of the non-local biphoton model of ghost imaging \cite{Scarcelli, Review2}. Our calculation indicates that $\langle\hat{E}^\dagger(\rho_i,z)\hat{E}(\rho_j,z) \rangle\propto\langle\hat{a}^\dagger(\kappa)\hat{a}(\kappa)\rangle$ when $\rho_i=\rho_j$, and is 0 otherwise. The correlation at any point on a single plane or paired points on CCD and bucket planes is proportional to the expectation value for a given mode in the momentum-basis. This behaviour reflects that of lens-based ghost imaging. In fact, it shows that we have a proportionality between photon counts in the source-plane and detection planes, which is not an entirely surprising result.

We can also go further and define effective operators in a similar manner to how we did for lens-based imaging by writing $\hat{E}_{\text{p}}=\frac{1}{\sqrt{A_p}}\int_{A_p} \hat{E}(\rho,z) d\rho$ where we are integrating over a pixel with area $A_p$. We can test that the correct commutation relation holds for this operator under the expectation value using similar assumptions as above.

It is then also easy to show that $\langle\hat{E}_{\text{p}}^\dagger\hat{E}_{\text{p}}\rangle\propto\langle\hat{a}^\dagger(\kappa)\hat{a}(\kappa)\rangle$. Again, the behaviour reflects that of lensed ghost imaging. This leaves an integration over the bucket detector to fully characterise our scheme. The only difference to the above calculation is that we are integrating over a larger area, $A_b$. It is evident that the auto-correlations will be the same. For the cross-correlations we find $\langle\hat{E}_{\text{b}}^\dagger\hat{E}_{\text{p}}\rangle\propto\langle\frac{\hat{a}^\dagger(\kappa)\hat{a}(\kappa)}{\sqrt{\frac{A_b}{A_p}}}\rangle$. Essentially, we scale the cross-correlations by the inverse square root of the ratio of the bucket detector area to the pixel area. This yields an identical scaling with changes in our three parameters $I, R, M$ as for the lens-based imaging.

\subsection*{Acknowledgments}
We  warmly acknowledge Adetunmise Dada for suggesting and motivating this work and Ladislav Mista Jr for discussions and early contributions. We further thank Matteo Paris, Madalin Guta, Ruggero Vasile and Davide Girolami for fruitful exchanges. We acknowledge the University of Nottingham for financial support through an Early Career Research and Knowledge Transfer Award and a Research Development Fund grant (EPSRC/RDF/BtG/0612b/31).

\subsection*{Contributions}
SR provided the leading contribution to the preparation, development, execution and interpretation of this work. GA contributed to the concept, interpretation and presentation of the results. Both authors contributed to writing the manuscript and preparing the illustrations.

\subsection*{Competing financial interests}
The authors declare no competing financial interests.

\subsection*{Corresponding author}
Correspondence to: Gerardo Adesso (\texttt{gerardo.adesso@nottingham.ac.uk})


\begin{thebibliography}{99}
\bibitem{Sergienko}
Strekalov, D. V.,  Sergienko, A. V., Klyshko, D. N. \& Shih, Y. H.
Observation of two-photon ``ghost'' interference and diffraction.
\textit{Phys. Rev. Lett.} \textbf{74}, 3600-3603 (1995)

\bibitem{book}
Kolobov, M. I. (editor) {\it Quantum Imaging}. (Springer, New York, 2007)

\bibitem{Pittman}
Pittman, T. B., Shih, Y. H., Strekalov, D. V. \& Sergienko A. V. Optical imaging by means of two-photon quantum entanglement, \textit{Phys. Rev. A} \textbf{58} R3429-R3432 (1995)

\bibitem{Abouraddy}
Abouraddy, A. F., Saleh B. E. A, Sergienko A. V., Teich M. C. Role of entanglement in two-photon imaging. \textit{Phys. Rev. Lett.} \textbf{87} 123602 (2001)

\bibitem{Bennink}
Bennink, R. S., Bentley, S. J. \& Boyd, R. W. ``Two-photon'' coincidence imaging with a classical source. \textit{Phys. Rev. Lett.} \textbf{89} 113601 (2002)

\bibitem{Gatti}
Gatti, A., Brambilla, B. \& Lugiato, L. A. Entangled imaging and wave-particle duality: from the microscopic
to the macroscopic realm. \textit{Phys. Rev. Lett.} \textbf{90} 113603 (2003)

\bibitem{Bennink2} Bennink, R. S., Bentley, S. J., Boyd, R. W. \& Howell, J. C. Quantum and classical coincidence imaging. \textit{Phys. Rev. Lett.} \textbf{92}, 033601 (2004)

\bibitem{Brambilla}
Brambilla E., Gatti A., Bache M. \& Lugiato L. A. Simultaneous near-field and far-field spatial quantum correlations in the high-gain regime of parametric down-conversion. \textit{Phys. Rev. A} \textbf{69} 023802 (2004)

\bibitem{Gatti2}
Gatti, A., Brambilla, E., Bache, M. \& Lugiato L. A. Ghost imaging with thermal light: comparing entanglement and classical correlation. \textit{Phys. Rev. Lett.} \textbf{93} 093602 (2004)

\bibitem{Valencia}
Valencia, A., Scarcelli, G., D'Angelo, M. \& Shih, Y. Two-photon imaging with thermal light. \textit{Phys. Rev. Lett.}  \textbf{94} 063601 (2005)

\bibitem{Ferri}
Ferri, F. et al. High-resolution ghost image and ghost diffraction experiments with thermal light. \textit{Phys. Rev. Lett.}  \textbf{94} 183602 (2005)

\bibitem{Debate1}
Shih, Y. H. The physics of ghost imaging -- nonlocal interference or local intensity fluctuation correlation? \textit{Quant. Inf. Proc.} DOI: 10.1007/s11128-012-0396-5 (2012)

\bibitem{Debate2}
Shapiro, J. H. \& Boyd, R. W. Response to ``The physics of ghost imaging -- nonlocal interference or local intensity fluctuation correlation?'' \textit{Quant. Inf. Proc.} DOI: 10.1007/s11128-012-0399-2 (2012)


\bibitem{Erkmen}
Erkmen, B. I. \& Shapiro, J. H. A unified theory of ghost imaging with Gaussian state light. \textit{Phys. Rev. A} \textbf{77} 043809 (2008)

\bibitem{ShapiroSNR} Erkmen, B. I. \&  Shapiro, J. H. Signal-to-noise ratio of Gaussian-state ghost imaging. \textit{Phys. Rev. A} \textbf{79}, 023833 (2009)

\bibitem{Sullivan}
O'Sullivan, M. N., Chan, K. W. C \&  Boyd, R. W. Comparison of the signal-to-noise characteristics of quantum versus thermal ghost imaging, \textit{Phys. Rev. A} \textbf{82}, 053803 (2010)

\bibitem{Review}
Shapiro, J. H. \& Boyd, R. W. The physics of ghost imaging. \textit{Quant. Inf. Proc.} DOI: 10.1007/s11128-011-0356-5 (2012)

\bibitem{Discord}
Adesso, G. \& Datta, A. Quantum versus classical correlations in Gaussian states.  \textit{Phys. Rev. Lett.} \textbf{105} 030501 (2010)

\bibitem{Discord2}
Giorda, P. \& Paris, M. G. A. Gaussian quantum discord. \textit{Phys. Rev. Lett.}
 \textbf{105}, 020503 (2010)

\bibitem{Zurek} Ollivier, H. \&  Zurek, W. H.
Quantum discord: a measure of the quantumness of correlations. \textit{Phys. Rev. Lett.} \textbf{88}, 017901 (2001)

\bibitem{Vedral} Henderson, L. \& Vedral, V. Classical, quantum and total correlations. \textit{J. Phys. A: Math. Gen.} \textbf{34}, 6899-6905 (2001)

\bibitem{Mandel}
Mandel L. \& Wolf E. \textit{Optical Coherence and Quantum Optics} (Cambridge Univ. Press, Cambridge, 1995)

 \bibitem{ourreview}
 Adesso, G. \& Illuminati, F. Entanglement in continuous-variable systems: recent advances and current perspectives. \textit{J. Phys. A: Math. Theor.} \textbf{40} 7821-7880, (2007)

\bibitem{slater}
Marian, P., Marian, T. A. \& Scutaru, H.
Inseparability of mixed two-mode Gaussian states generated with a SU(1,1) interferometer.
\textit{J. Phys. A: Math. Gen.} \textbf{34}, 6969-6980 (2001)

\bibitem{ferraro}
Ferraro, A., Aolita, L., Cavalcanti, D., Cucchietti, F. M. \& Ac\'in, A. Almost all quantum states have nonclassical correlations. \textit{Phys. Rev. A} \textbf{81}, 052318 (2010)

\bibitem{dreview}
Modi, K., Brodutch, A., Cable, H.,  Paterek, T. \& Vedral, V.
Quantum discord and other measures of quantum correlation. \textit{e--print} arXiv:1112.6238v1 (2011)

\bibitem{merali} Merali, Z. Quantum computing: the power of discord. \textit{Nature} \textbf{474}, 24-26 (2011)

\bibitem{Matteo}
Ferraro A. \& Paris, M. G. A. Non-classicality criteria from phase-space representations and information-theoretical constraints are maximally inequivalent. {\it Phys. Rev. Lett.} \textbf{108} 260403 (2012)

\bibitem{Scarcelli}
Scarcelli, G., Berardi, V. \& Shih, Y. H. Can two-photon correlation of chaotic light be considered as correlation of intensity fluctuations? \textit{Phys. Rev. Lett.} \textbf{96} 063602 (2006)

\bibitem{Sun}
Karmakar, S. \& Shih, Y. The first experimental demonstration of ghost imaging with sunlight. \textit{Proc. SPIE} \textbf{8400} (2012)


\bibitem{Turb1}
Meyers, R. E., Deacon K. S. \& Shih Y. Turbulence-free ghost imaging \textit{Appl. Phys. Lett.} \textbf{98} 111115 (2011)

\bibitem{Turb2}
Shapiro, J. H. Comment on ``Turbulence-free ghost imaging'' \textit{e--print} arXiv:1201.4513v1 (2012)

\bibitem{Review2}
Shih, Y. The physics of ghost imaging \textit{Advances in Lasers and Electro Optics} DOI: 10.5772/8663 (2010)


\bibitem{Speckle}
Goodman, J. W. \textit{Speckle Phenomena in Optics: Theory and Applications} (Roberts, Greenwood Village, Colorado, 2007)

\bibitem{Brida}
Brida, G. et al. Systematic analysis of signal-to-noise ratio in bipartite ghost imaging with classical and quantum light. \textit{Phys. Rev. A} \textbf{83} 063807 (2011)


\bibitem{ulrik}
Madsen, L. S., Berni, A., Lassen, M. \& Andersen, U. L. Experimental Investigation of the Evolution of Gaussian Quantum Discord in an Open System. \textit{Phys. Rev. Lett.} {\bf 109} 030402 (2012)

\bibitem{milegu}
Gu, M. et al. Observing the operational significance of discord consumption. {\it Nature Phys.} doi:10.1038/nphys2376 (2012) 


\bibitem{blandino}
Blandino, R. et al. Homodyne estimation of Gaussian quantum discord. \textit{e--print} arXiv:1203.1127v2 (2012)

\bibitem{Rubin}
Rubin, M. H. Transverse correlation in optical spontaneous parametric down-conversion. \textit{Phys. Rev. A} \textbf{54} 5349-5360 (1996)

\bibitem{Shapiro}
Yuen, H. P. \& Shapiro, J. H. Optical communication with two-photon coherent states -- part I: quantum-state propagation and quantum-noise reduction. \textit{IEEE Trans. Inform. Theory} \textbf{24} 657-668 (1978)

\end{thebibliography}
\end{document}